\documentclass[aps,prc,groupedaddress,showpacs,showkeys]{revtex4-1}
\usepackage{latexsym,epsfig,amsmath,amssymb}
\usepackage{graphicx}
\usepackage{graphics}

\newcommand{\seq}{\begin{subequations}}
\newcommand{\sen}{\end{subequations}}
\newcommand{\eq}{\begin{eqnarray}}
\newcommand{\en}{\end{eqnarray}}

\def\L2{\Lambda^2}

\begin{document}

\title{Deuteron electromagnetic form factors in transverse plane
with a phenomenological Lagrangian approach}
\author{Cuiying Liang$^{1,2,3}$, Yubing Dong$^{1,2}$, and Weihong Liang$^{3}$
\vspace*{.3\baselineskip}\\}
\affiliation{
$^{1}$ Institute of High Energy Physics, Chinese Academy of Sciences,
Beijing 100049, P. R. China \vspace*{.3\baselineskip}\\
$^{2}$ Theoretical Physics Center for Science Facilities (TPCSF),
CAS, P. R. China \vspace*{.3\baselineskip}\\
and \vspace*{.3\baselineskip}\\
$^{3}$ College of Physical Science and Technology, Guangxi Normal University,
 Guilin 541004, P. R. China\\}
\date{\today}

\begin{abstract}

A phenomenological Lagrangian approach is employed to study the
electromagnetic properties of deuteron. The deuteron is regarded as
a loosely bound state of a proton and a neutron. The deuteron
electromagnetic form factors are expressed in light-front
representation in the transverse plane. The transverse charge
density of the deuteron is discussed.
\end{abstract}



\maketitle
%\newpage
\section{Introduction}
It is known that the study of electromagnetic(EM) form factors of proton,
neutron and light nuclei, like deuteron and He-3, is crucial for the
understanding of nucleon structures. It tells the distributions of the
charge and magnetization inside systems.  The EM form factors
of the deuteron have been explicitly discussed (for some recent reviews, see,
e.g.~\cite{Gilman}-\cite{Garcon}) for several decades. A deuteron,
as a spin-1 particle, has three form factors of charge $G_C$, magnetic $G_M$,
and quadrupole $G_Q$.  It is often regarded as a loosely
bound state of the proton and neutron (with binding energy
$\epsilon_D \sim 2.22$~MeV), and consequently
the study of the deuteron properties can shed light on the structure of the
nucleon as well as nuclear effects. Moreover, it is found that the two
constituents -- proton and neutron inside the deuteron are dominated by
the relative $S$-wave, and the $D$-wave is only about $5\%$. The understanding
of the deuteron structures, like its EM form factors and its binding energy,
is usually based on potential models, on phenomenological
models with quark, meson, and nucleon degrees of freedom, and on some
effective field theories etc. \cite{Gilman, Sick, Gross, Garcon, Arnold,
Mathiot, Wiringa, Arhen, Gari, Karm, Kaplan, Ivanov}. The realistic deuteron
wave function has already been explicitly given by Ref. \cite {Mau},
particularly, the relativistic deuteron wave function was discussed and
obtained in Refs. \cite {Gross1, Carbonell, Strikman}. \\

\begin{figure}
  \centering
  \includegraphics[width=6cm, height=6cm]{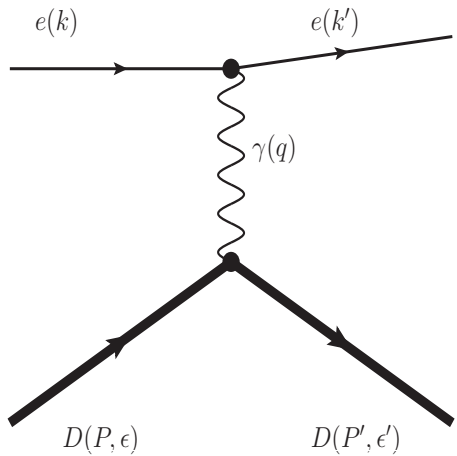}\\
  \caption{Feynman diagram for electron-deuteron elastic scattering
  in the one-photon approximation.}
\end{figure}

Recently, the pion transverse charge density $\rho_C(b)$ is of great interests.
It stands for the two-dimensional Fourier transform of the EM form factor and
for the density (in the infinite momentum frame) located at a transverse
separation $b$ (impact parameter) from the center of transverse momentum
\cite {Soper,Burkardt, Diehl, Carlson, Miller0}. It is pointed out that
this two-dimensional density can directly relate to the matrix element of a
density operator. However, the usual three-dimensional Fourier transforms
of the form factors cannot, since the initial and final momentums are
different and one cannot boost the initial and final states to the rest
frame simultaneously.  There are also many discussions on the proton EM form
factors in the transverse plane. \\

Analogous to pion, in this paper, we will study the EM form factors of the
deuteron in the transverse plane. A phenomenological approach will be employed
for the deuteron, where it is regarded as a loosely bound state of a proton
and a neutron and the two constituents are in relative $S$-wave. The coupling
of the deuteron to its two composite particles is determined by the known
compositeness condition from Weinberg \cite {Weinberg}, Salam \cite {Salam}
and  others \cite {Hay, Efi}. Our approach has been successfully applied to
study the properties of weakly bound state problems, like the new resonances
of $X(3872)$, $\Lambda_c(2940)$, and the EM form factors of pion as well as
some other observables \cite{Dong0, Dong1}.\\

This paper is organized as follows. In section II, the general properties
of the deuteron (spin-1 particle) is briefly reviewed,  and moreover, our
phenomenological approach is briefly explained. In section III, the EM form
factors of the deuteron in the light-front representation are given. Our
numerical results for the EM form factors in the transverse plane are
shown in section IV.  Finally, section V is devoted for a short summary.

\section{Our framework}

\subsection{Deuteron electromagnetic form factors}

A deuteron is a spin-1 particle, and its EM properties can be explored by
a lepton-deuteron  elastic scattering. The matrix element for
electron-deuteron ($eD$) elastic scattering in the one-photon approximation,
as shown in Fig. 1, can be written as
\eq
{\cal M}=\frac{e^2}{Q^2} \bar u_e(k^\prime) \gamma_{\mu}
u_e(k) {\cal J}_{\mu}^D(P,P^\prime),
\en
where $k$ and $k^\prime$ are the four--momenta of initial and final
electrons. ${\cal J}_{\mu}^D(P,P^\prime)$ is the deuteron EM current, and its
general form is
\eq\label{D_current}
{\cal J}_{\mu}^D(P,P^\prime) =
- \biggl( G_1(Q^2)\epsilon^{\prime *}\cdot\epsilon-\frac{G_3(Q^2)}{2M_D^2}
\epsilon\cdot q\epsilon^{\prime *}\cdot q \biggr) (P + P^\prime)_{\mu}
- G_2(Q^2) \biggl( \epsilon_{\mu}\epsilon^{\prime *}\cdot q
- \epsilon^{\prime *}_{\mu} \epsilon\cdot q \biggr)~,
\en
where $M_D$ is the deuteron mass, $\epsilon$($\epsilon^\prime$) and
$P(P^\prime)$ are polarization and four--momentum of the initial (final)
deuteron,  and $Q^2=-q^2$ is momentum transfer square with $q=P^\prime - P$.
The three EM form factors $G_{1,2,3}$ of the deuteron
are related to the charge $G_C$, magnetic $G_M$, and quadrupole $G_Q$
form factors by
\eq
G_C = G_1+\frac23\tau G_Q\,, \hspace*{.25cm}
G_M \ = \ G_2 \,,            \hspace*{.25cm}
G_Q = G_1-G_2+(1+\tau)G_3,   \hspace*{.25cm}
\en
with $\tau=\frac{Q^2}{4M_D^2}$. The three form factors are normalized at
zero recoil as
\eq
G_C(0)=1\,, \ \ \
G_Q(0)=M_D^2{\cal Q}_D=25.83\,, \ \ \
G_M(0)=\frac{M_D}{M_N}\mu_D=1.714 \, ,
\en
where $M_N$ is the nucleon mass,
${\cal Q}_D$ and $\mu_D$ are the quadrupole and magnetic moments
of the deuteron.\\

The unpolarized differential cross section for the $eD$ elastic
scattering can be expressed by the two structure functions, $A(Q^2)$ and
$B(Q^2)$, as
\begin{equation}
{{d\sigma} \over {d\Omega}} = \sigma_M \left[ A(Q^2) +
B(Q^2) \tan^2{\left(\frac{\theta}{2}\right)} \right ] ,
\label{eq:eq10}
\end{equation}
where $\sigma_M=\alpha^2 E^\prime \cos^2(\theta/2)/[4 E^3 \sin^4(\theta/2)]$
is the Mott cross section for point-like particle, $E$ and $E^\prime$
are the incident and final electron energies, $\theta$ is the electron
scattering angle, $Q^2=-q^2=4 E E^\prime \sin^2(\theta/2)$,
and $\alpha=e^2/4\pi=1/137$ is the fine-structure constant.
The two form factors $A(Q^2)$ and $B(Q^2)$  are
related to the three EM form factors of the deuteron as
\begin{equation}
{ A(Q^2) =  { G^2_C(Q^2) + {8 \over 9} \tau^2 G^2_Q(Q^2) +
{2 \over 3} \tau G^2_M(Q^2) }  },~~~~~
{ B(Q^2) =  {4 \over 3} \tau (1+\tau) G^2_M(Q^2) }.
\label{eq:eq11}
\end{equation}
Clearly, the three form factors $G_{C,M,Q}$ cannot be simply determined by
measuring the unpolarized elastic $eD$ differential cross section.
To uniquely determine the three form factors of the deuteron one additional
polarization variable is necessary. For example, one may take the polarization
of $T_{20}$ \cite{Garcon}
\eq
T_{20}=-\displaystyle\frac{1}{\sqrt{2}{\cal S}}\left \{
\displaystyle\frac{8}{3}\tau G_C G_Q + \displaystyle\frac{8}{9} \tau^2
G_Q^2+\frac{1}{3} \tau
\left [1+2(1+\tau)\tan^2 (\theta /2)\right ]G_M^2\right \},
\en
into account, where ${\cal S}=A+B\tan^2 (\theta /2)$.\\

\subsection{The phenomenological approach}

\begin{figure}
  \centering
  \includegraphics[width=12cm]{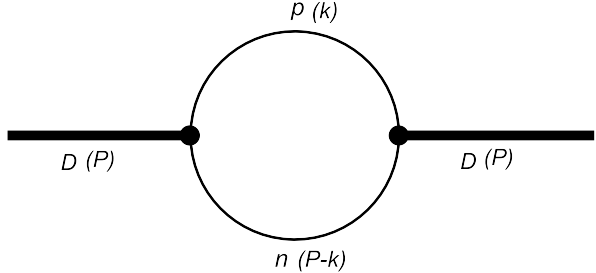}\\
  \caption{Deuteron mass operator}
\end{figure}

Here we will briefly show the formalisms of the phenomenological
approach. Take an assumption that the deuteron is interpreted as a
hadronic molecule -- a weakly bound state of the proton and neutron:
$ |D\rangle=|pn\rangle$ (see Fig. 2), then one may simply write a
phenomenological effective Lagrangian of the deuteron and its two
constituents -- proton and neutron, as \eq {\cal
L}_D(x)=g_DD^{\dagger}_{\mu}(x)\int
dy\bar{p}^c(x+y/2)\widetilde{\Phi}_D (y^2)\Gamma^{\mu}n(x-y/2)+H.c.,
\en where $D_{\mu}$ is the deuteron field, $\bar{p}^c(x)=p^T(x)C$,
$C$ denotes the matrix of charge conjugation, and $x$ is the
centre-of-mass (C. M.) coordinate. In Eq. (8),
$\widetilde{\Phi}_D(y^2)\Gamma^{\mu}$ is the vertex where the
correlation function $\widetilde{\Phi}_D(y^2)$ characterizes the
finite size of the deuteron as a $pn$ bound state and
depends on the relative Jacobi coordinate $y$. \\

A basic requirement for the choice of an explicit form of this correlation
function is that its Fourier transform vanishes sufficiently fast in the
ultraviolet region of Euclidean space  to render the Feynman diagrams
ultraviolet finite. Usually a Gaussian-type function is selected as
the correlation for simplicity. One chooses
\eq {\tilde
\Phi}_D(k^2)\Gamma^{\mu}=\exp(-k_E^2/\Lambda_D^2)\gamma^{\mu},
\en
for the Fourier transform of the correlation function, where
$\Gamma^{\mu}=\gamma^{\mu}$, $k_E$ is the Euclidean Jacobi momentum
and $\Lambda_D$ is a free size parameter which represents the distribution
of the two constituents in the deuteron.\\

The coupling of $g_D$ in Eq. (8) can be determined by the known
compositeness condition, which implies the renormalization constant
of the hadron wave function is set equal to zero as
$Z_D=1-\Sigma_D'(M_D^2)=0$, with
$\Sigma_D'(M_D^2)=g_D^2\Sigma_{D\perp}'(M_D^2)$
being the derivative
of the transverse part of the mass operator (see Fig. 2). Usually,
the mass operator splits into the transverse part $\Sigma_{D\perp}(k^2)$
and longitudinal one $\Sigma_{D\parallel}(k^2)$ as
\eq
\Sigma^{\alpha\beta}_D(k)=g^{\alpha\beta}_{\perp}\Sigma_{D\perp}(k^2)+
\frac{k^{\alpha}k^{\beta}}{k^2}\Sigma_{D\parallel}(k^2),
\en
where $g_{\perp}^{\alpha\beta}=g^{\alpha\beta}-k^{\alpha}k^{\beta}/k^2$
and $g_{\perp}^{\alpha\beta}k_{\alpha}=0$.  From Eqs. (9-10) we see that
for a fixed parameter $\Lambda_D$, the coupling of the deuteron to its
constituents -- proton and neutron, $g_D$, is well determined by
the compositeness condition. The explicit expression of $g_D$ (in the
simplest case of Eq. (8)), in terms of the loop integral shown in Fig. 2, has
been given in Refs. \cite{Dong1, Dong2}. The correlation
function of Eq. (9) simulates only the $S$-wave in the deuteron. It is
commonly believed that the relative $S$-wave is dominant in the deuteron.

\section{The light-front representation}

\subsection{EM form factors in the light-front representation}

To study the EM properties of the deuteron, we assume the deuteron as
a bound state of the proton and neutron. Therefore, the $eD$
scattering can be interpreted as the photon coupling respectively to the
proton and neutron, as shown in Fig. 3. The general expression of the
loop integral of Fig. 3 is
\begin{figure}
\centering
\includegraphics[width=12cm]{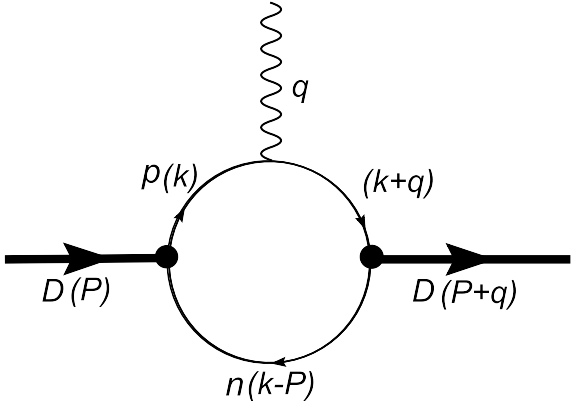}\\
\caption{Electron-deuteron scattering diagram contributing to
the EM form factors.}
\end{figure}

\eq
i{\cal M}^{\alpha}=i\epsilon^*_{\mu}{\cal M}^{\alpha\mu\nu}\epsilon_{\nu}
\en
where
\eq
{\cal M}^{\alpha\mu\nu}&&=-ig_D^2\int {d^4k\over (2\pi)^4}
\frac{Tr{[\gamma^{\mu}(\slash \!\!\! k+\slash \!\!\! q+M_N)
\Gamma^{\alpha}(\slash \!\!\! k+M_N)\gamma^{\nu}
(\slash \!\!\! k-\slash \!\!\! P+M_N)]}}{(k^2-M_N^2)[(k+q)^2-M_N^2]
[(k-P)^2-M_N^2]}\\ \nonumber
&&\times \tilde\Phi_D\big ((k-P/2)_E^{2}\big )
\tilde\Phi_D\big ((k-P/2+q/2)_E^{2}\big ),
\en
and
\eq
\Gamma^{\alpha}=\gamma^{\alpha}[F_1^p(Q^2)+F_1^n(Q^2)]
+i\frac{\sigma^{\alpha\beta} q_{\beta}}{2M_N}[F_2^p(Q^2)+F_2^n(Q^2)],
\en
where $F_{1,2}^{p,n}$ are the Dirac and Pauli form factors of the proton and
neutron, respectively. In Eq. (12), ${\tilde \Phi}_D$ stands for the
correlation function.\\

With the help of the calculation of the scalar loop integral of
$I(P^2,q^2,P\cdot q)$ shown in appendix, one may easily compute the matrix
element ${\cal M}^{\alpha\mu\nu}$ in Eq. (12) in the light-front
representation. Furthermore, one may obtain the model-dependent deuteron form
factors according to the general Lorentz structure given by Eq. (2).
Taking the charge form factor of the deuteron for example, the obtained
form factor is
\eq
G_C(Q^2)&=&g_D^2\sum_{N=p,n}
\int {dxd^2\vec{k}\over (2\pi)^3}\frac{2
\big [M_D^2(1+x)+xQ^2+\vec{k}\cdot \vec{q}\big ]
F_1^N(Q^2)-Q^2F_2^N(Q^2)}{x^2(1-x)\Big [P^+P^--\frac{\vec{k}^2+M_N^2}{x}
-\frac{(\vec{k}-\vec{P})^2+M_N^2}{1-x}\Big ]
\Big [P^+P^--\frac{(\vec{k}+\vec{q})^2+M_N^2}{x}
-\frac{(\vec{k}-\vec{P})^2+M_N^2}{1-x}\Big ]}\\ \nonumber
&&\times \exp\Bigg [-\frac{1}{\Lambda_D^2}(x-\frac{1}{2})\Big [\frac{P^+P^-}
{2}-\frac{(\vec{k}-\frac{\vec{P}}{2})^2}{x-\frac{1}{2}}
-\frac{(\vec{k}-\vec{P})^2+M^2_N}{1-x}\Big ]\Bigg ]\\ \nonumber
&&\times \exp\Bigg [-\frac{1}{\Lambda_D^2}(x-\frac{1}{2})\Big [\frac{P^+P^-}
{2}-\frac{(\vec{k}-\frac{\vec{P}}{2}+\frac{\vec{q}}{2})^2}{x
-\frac{1}{2}}-\frac{(\vec{k}-\vec{P})^2+M^2_N}{1-x}\Big ]\Bigg ],
\en
where $P^{\pm}=P^0\pm P^3$, and $x=k^+/P^+$. In addition, we define the
transverse momentum
\begin{equation}
\vec{\kappa}=(1-x)\vec{k}-x(\vec{P}-\vec{k})=\vec{k}-x\vec{P},
\end{equation}
then the charge form factor can be re-written as
\begin{eqnarray}
G_C(Q^2)&=&g_D^2\sum_{N=p,n}
\int {dxd^2\vec{\kappa}\over{(2\pi)^3}}
\frac{\big [2M_D^2(1+x)+\vec{\kappa}\cdot \vec{q}\big ]
F_1^N(Q^2)-Q^2F_2^N(Q^2)}
{x^2(1-x)\Big [M_D^2-\frac{\vec{\kappa}^2+M_N^2}{(1-x)x}\Big ]\Big [M_D^2
-\frac{[\vec{\kappa}+(1-x)\vec{q}]^2+M_N^2}{(1-x)x}\Big ]}\\ \nonumber
&&\times \exp\Bigg [\frac{1}{\Lambda_D^2}(x-\frac{1}{2})
\Big (\frac{M_D^2}{2}
-\frac{M_N^2}{1-x}-\frac{\vec{\kappa}^2}{2(x-\frac{1}{2})(1-x)}\Big )\Bigg ]
\\ \nonumber
&&\times \exp\Bigg [\frac{1}{\Lambda_D^2}
(x-\frac{1}{2})\Big (\frac{M_D^2}{2}-\frac{M_N^2}{1-x}
-\frac{[\vec{\kappa}+(1-x)\vec{q}]^2}{2(x-\frac{1}{2})(1-x)}\Big )\Bigg ].
\end{eqnarray}
Let us define a wave function $\psi$ as
\eq
\psi(x,\vec{\kappa})=\frac{1}{M_D^2-\frac{\vec{\kappa}^2+M_N^2}{(1-x)x}}
\exp\Bigg [\frac{1}{\Lambda_D^2}(x-\frac{1}{2})
\Big (\frac{M_D^2}{2}-\frac{M_N^2}{1-x}-\frac{\vec{\kappa}^2}
{2(x-\frac{1}{2})(1-x)}\Big ) \Bigg ],
\en
and finally the charge form factor is
\eq
G_C(Q^2)&=&g_D^2\sum_{N=p,n}
\int {dxd^2\vec{\kappa}\over (2\pi)^3x^2(1-x)}
\Bigg \{\Big [2M_D^2(1+x)+\vec{\kappa}\cdot \vec{q}\Big ]F_1^N(Q^2)
-Q^2F_2^N(Q^2)\Bigg \}
\psi(x,\vec{\kappa})\\ \nonumber
&&\times\psi^*(x,\vec{\kappa}+(1-x)\vec{q}).
\en
In the same way, the magnetic form factor is
\eq
G_M(Q^2)&=&g_D^2\sum_{N=p,n}
\int {dxd^2\vec{\kappa}\over (2\pi)^3x^2(1-x)}
\Bigg \{\Big [2M_D^2(1+3x)-6\frac{\vec{\kappa}^2+M_N^2}{1-x}+8M_N^2\Big]
F_1^N(Q^2)\\ \nonumber
&& -[2M_D^2(1+2x)-4\frac{\vec{\kappa}^2+M_N^2}{1-x}+4M_N^2\Big]F_2^N(Q^2)
\Bigg \}\psi(x,\vec{\kappa})\psi^*(x,\vec{\kappa}+(1-x)\vec{q}).
\en

\subsection{Electromagnetic form factors and transverse densities}

So far the form factor $G_C(Q^2)$ is expressed by a three-dimensional
integration that involves wave functions in  momentum-space (see
Eqs. (17-19)). The two-dimensional Fourier transform of the wave function
of Eq. (17) can be expressed as
\begin{eqnarray}
\psi(x,\textbf{B})&=& \frac{1}{\sqrt{(1-x)x^2}}
\int {d^2\vec{\kappa}\over (2\pi)^2}\psi(x,\vec{\kappa})
e^{i\vec{\kappa}\cdot\textbf{B}}\\ \nonumber
&&=\frac{\sqrt{1-x}}{2\pi}
\int_0^\infty dt\frac{\cos|B|t}{\sqrt{t^2+c^2}}
\Big (1-\Phi\big [\sqrt{\frac{t^2+c^2}{2\Lambda_D^2(1-x)}}\big ]\Big )
\exp\Big [\frac{1}{\Lambda_D^2}(M_N^2-\frac{M_D^2}{4})\Big ],
\end{eqnarray}
where $\Phi(x)$ is error function and $c^2=M^2_N-M^2_D(1-x)x$.
Thus, the charge form factor of Eq. (18) is expressed as
\begin{equation}
G_C(Q^2)=g_D^2\sum_{N=p,n}
\int_0^1dx\int {d^2\textbf{B}\over(2\pi)^3}
\Bigg \{2\big [M_D^2(1+x)-(1-x)Q^2\big ]F_1^N(Q^2)
-Q^2F_2^N(Q^2)\Bigg \}
\mid \psi(x,B)\mid ^2e^{-i\vec{q}\cdot(1-x)\textbf{B}}.
\end{equation}
In order to further simplify the expression of $G_C(Q^2)$, the relative
transverse position variable \textbf{B} is expressed by the value of
$\textbf{b}_1=\textbf{b}$, which is the transverse position variable
of the charged parton of the deuteron. We have
\eq
\textbf{B}=\textbf{b}_1-\textbf{b}_2=\frac{\textbf{b}}{1-x},
~~~~~\textbf{b}_1(x)+\textbf{b}_2(1-x)=0.
\en
With the help of Eqs. (21-22), we find
\begin{eqnarray}
G_C(Q^2)=\int_0^1{g_D^2dx\over{1-x}}\int {d^2\textbf{b}\over(2\pi)^3}
\sum_{N=p,n} \Bigg \{2\big [M_D^2(1+x)-(1-x)Q^2
\big ]F_1^N(Q^2)-Q^2F_2^N(Q^2)\Bigg \}
  \mid \psi(x,\frac{b}{1-x}) \mid^2 e^{-i\vec{q}.
\cdot\textbf{b}}.
\end{eqnarray}
The Fourier transform of the charge form factor is
\begin{equation}
G_C(Q^2)=\frac{1}{(2\pi)^2}\int {d^2\textbf{b}}\rho_C(b)
e^{-i\vec{q}\cdot \textbf{b}},
\end{equation}
where the quantity $\rho_C(b)$ stands for the transverse charge density of
the deuteron and $b$ is the impact parameter in the transverse
plane. Similarly, one may also determine the transverse magnetic density
in the transverse plane as
\eq
G_M(Q^2)=\frac{1}{(2\pi)^2}\int {d^2\textbf{b}}\rho_M(b)
e^{-i\vec{q}\cdot \textbf{b}},
\en
where quantity $\rho_M(b)$ is expressed in terms of
the impact parameter $b$.\\

\section{Numerical results}

After integrating over $x$ and $\textbf{b}$, we may estimate the obtained
$G_C(Q^2)$ and $G_M(Q^2)$. In our numerical calculations, the only
one model-dependent parameter is $\Lambda_D$ and we select
$\Lambda_D\sim 0.23GeV$\cite{Bunden}. In Figs. 4 and 5 our numerical results
for the charge and magnetic form factors of the deuteron are shown.
In order to compare our results with the experimental measurement, the
phenomenological parameterizations \cite{Tomasi} of the measured
two form factors are shown by solid lines in the two figures.\\

\begin{figure}
\centering
\includegraphics[width=9 cm]{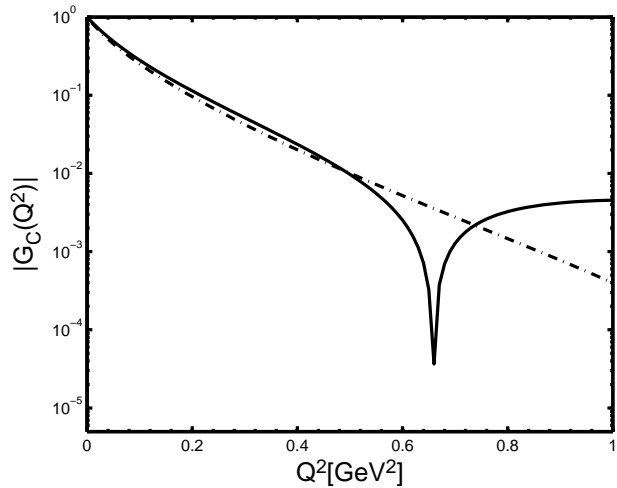}\\
\caption{ Form factor $|G_C(Q^2)|$. The solid line is from the
parameterizations of Ref. \cite{Tomasi} and the dash-dotted line is our result
in the light-front representation.}\label{55}
\end{figure}

\begin{figure}
\centering
\includegraphics[width=9cm]{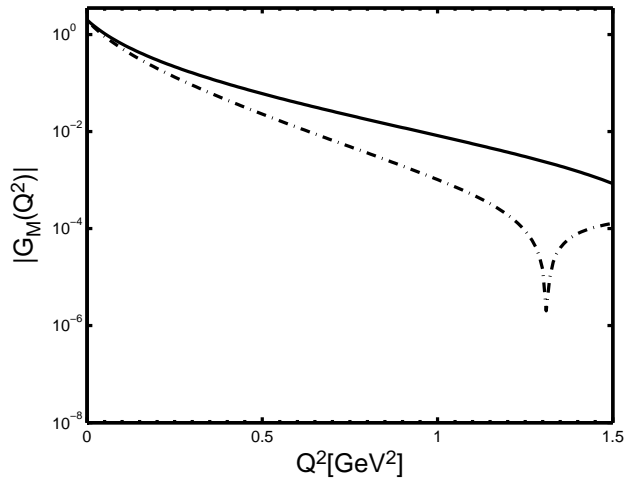}\\
\caption{Form factor $|G_M(Q^2)|$. The solid line is from the
parameterizations of Ref. \cite{Tomasi} and the dash-dotted line is our result
in the light-front representation}
\end{figure}

Figs. 4 and 5 show that the present approach could, at least qualitatively,
reproduce the EM form factors of the deuteron in the low $Q^2$ region,
although there are discrepancies between our results and the parameterized
form factors. The discrepancies become larger when the $Q^2$ increases.
This phenomenon is not surprising. This is due to the fact that our present
approach is rather simple and we have only one parameter $\Lambda_D$.
Moreover, we have employed the Gauss-like correlation function of Eq. (9)
to simplify the deuteron wave function for the calculation. However the
Guass-like wave function usually drops faster than the realistic
case \cite{Tomasi}. It should be mentioned that in order to get the best
fits for the deuteron EM form factors, 4 free parameters are employed for
each of the three form factors in the parameterization scheme of
Ref. \cite{Tomasi}, just as it claimed that the dipole electric and magnetic
form factors of the proton and neutron as well as the meson cloud
effect (of $\rho$ and $\omega$ mesons) are considered simultaneously
in \cite{Tomasi}. There are several sets of the parameterization in Ref.
   \cite{Tomasi}. They can reproduce the data in the low $Q^2$ region
quite well.\\

The important quantities of the present calculation are the transverse charge
and magnetic densities $\rho_{C,M}(b)$ of the deuteron. They are written in
terms of the impact parameter $b$ in the transverse plane and they stand
for the charge and magnetic densities of the deuteron in the transverse plane.
In Figs. 6 and 7, we plot the estimated $\rho_{C,M}(b)$ comparing to the
results from the parameterized form factors, where the red solid curves and
the black dashed curves are obtained from the two-dimensional Fourier
transform of the parameterized charge and magnetic form factors \cite{Tomasi}
as
\begin{eqnarray}
\rho_{C,M}(b)&=&\int{d^2q\over(2\pi)^2}G_{C,M}(\vec{q}^{~2})
e^{i\vec{q}\cdot \vec{b}}\nonumber\\
       &=&\int_0^\infty{qdq\over2\pi} G_{C,M}(\vec{q}^{~2})J_0(qb),
\end{eqnarray}
with $J_0(qb)$ being a cylindrical Bessel function and $Q^2=-q^2=\vec{q}^{~2}$.
In Fig. 8 we plot the transverse
charge density $\rho_C(b)$ for the proton and neutron for an example. \\

\begin{figure}
\centering
\includegraphics[width=12cm]{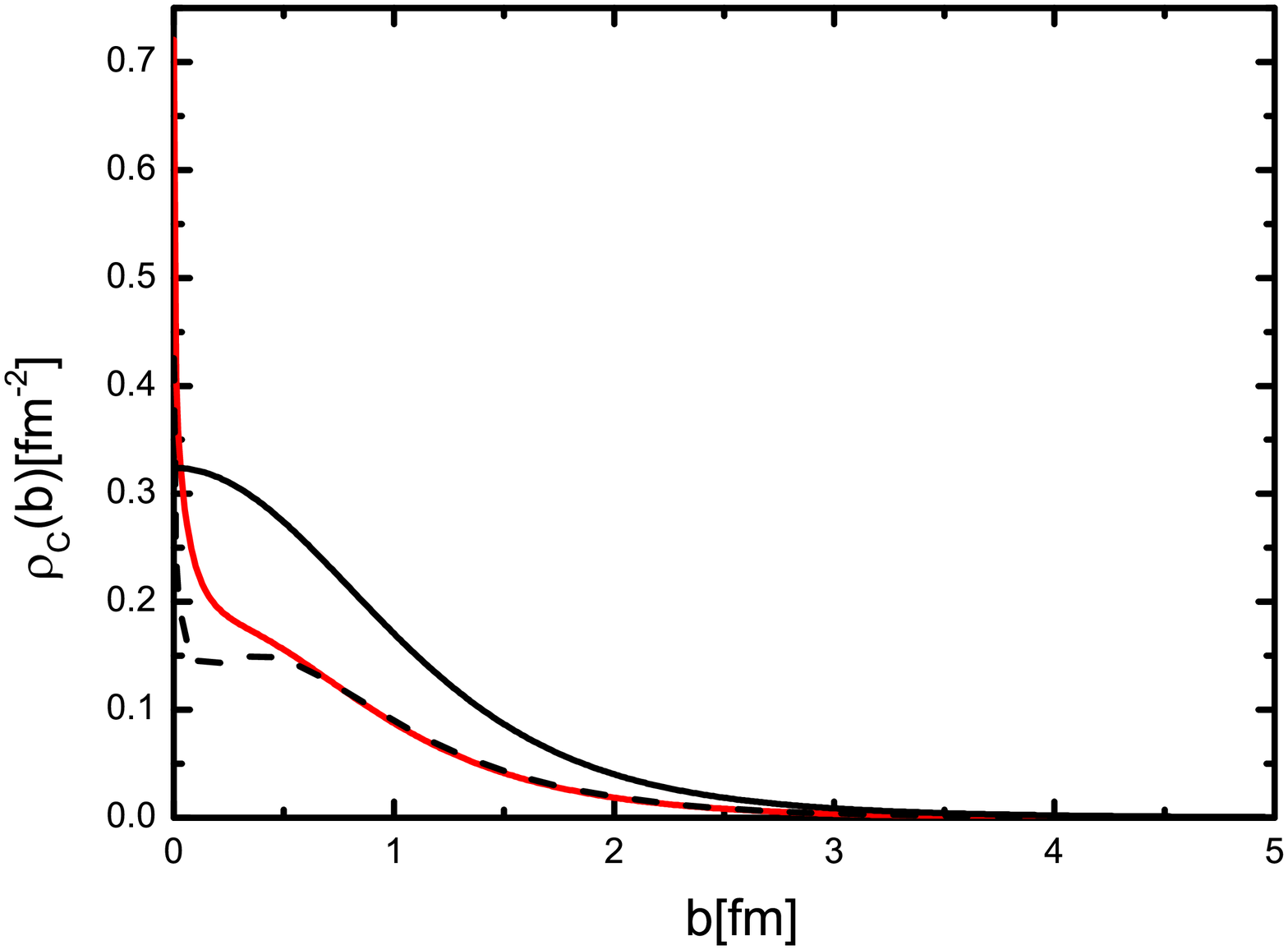}\\
\caption{The transverse charge density of deuteron. The red solid
line and the black dashed line are the results of the
parameterizations, the black solid line is our result in the
light-front representation.}
\end{figure}

\vspace{0.5cm}

\begin{figure}
\centering
\includegraphics[width=12cm]{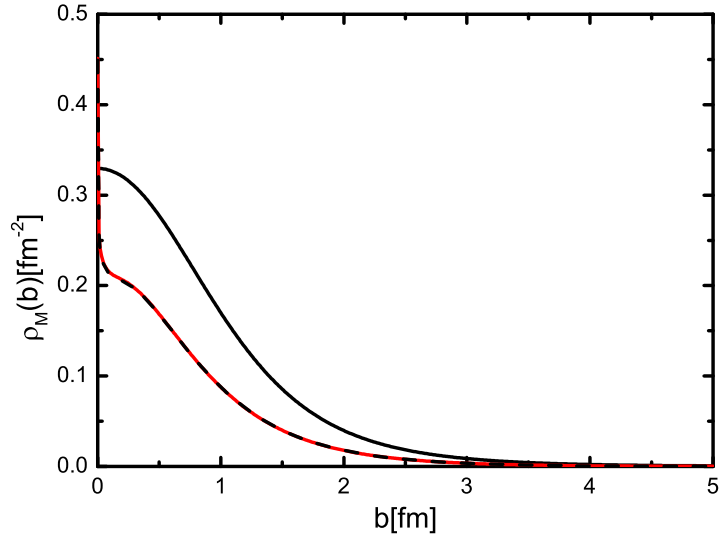}\\
\caption{The transverse magnetic density of deuteron. The red solid
line and the black dashed line are the results of the
parameterizations, the black solid line is our result in the
light-front representation.}
\end{figure}

From Fig. 6, one can see that our result fits the one of the parametrization only
qualitatively, and one can also find that different parametrizations have
different results at $b=0$ (see red solid and black dashed curves). The red
solid line (black dashed line) gets the maximum value 0.720 (0.426) at
$b=0$, and our black solid line gets the maximum value 0.324. The remarkable
differences are because the parametrizations are from the fit to the
experimental data in the low $Q^2$ region, which corresponds to large $b$
region. Therefore, for the small $b$ region, the uncertainty is expected to
be large since our knowledge for the form factors in the large $Q^2$ region
is limited. \\

We can determine the deuteron mean-square transverse charge
radius. It is defined as \eq \langle b_C^2\rangle=\int d^2b b^2\rho_C(b), \en and
it can also be yielded from \eq \lim_{Q^2\to
0}G_C(Q^2)=1-\frac{Q^2}{4}\langle b_C^2\rangle. \en It stands for the size of the
deuteron in the transverse plane. This quantity differs from the
well-known effective mean-square charge radius $R^{*2}$ in the
three-dimension space of \eq \lim_{Q^2\to
0}G_C(Q^2)=1-\frac{Q^2}{6}R_C^{*2}. \en The relation of the two
quantities is $\langle b_C^2\rangle=\frac23 R_C^{*2}$ \cite{Miller2}. In our
calculation, we obtained $R_C^*\sim 2.75fm$, which is consistent
with the value of $R_C^*\sim 2.56fm$ from the parameterizations
\cite{Tomasi} and the experimental extraction of $R_C^*=2.128\pm
0.11fm$ \cite{expt}. The magnetic radius we obtained is about
$R^*_M\sim 2.16~fm$ which also reasonably fits the result from the
parameterizations of $R_{M}^*\sim 1.93~fm$ and the experimental data
of $R_{M}^*=1.90\pm 0.14~fm$ \cite{expt}. \\

\begin{figure}
  \centering
  \includegraphics[width=12 cm]{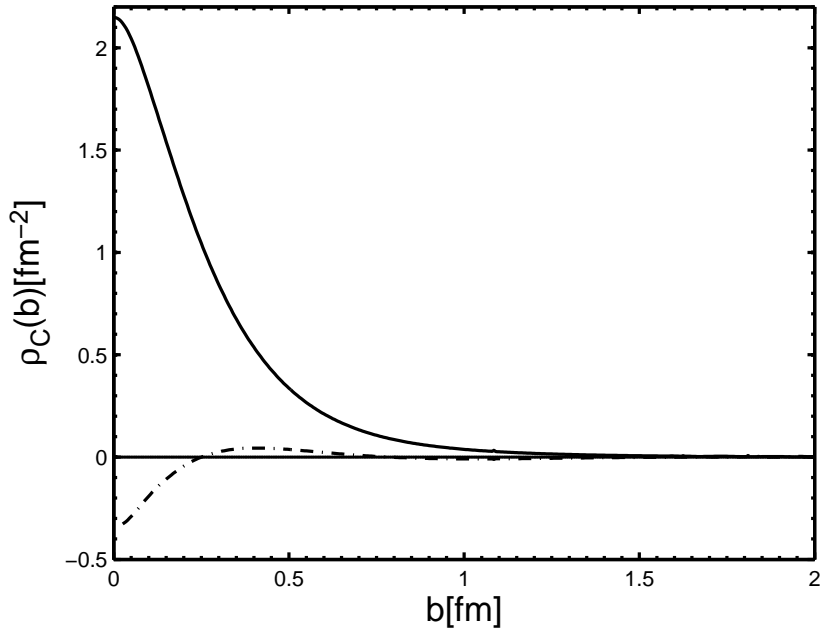}\\
  \caption{The transverse charge density of proton and neutron. The
solid line is the transverse charge density of proton and the
dotted-dashed line is the transverse charge density of neutron.}
\end{figure}

The obtained $\rho_{C,M}(b)$, in Figs. 6 and 7, tell that they have
similar $b$-dependences to the parameterized ones. Although the
discrepancies exist in the small $b$ region which corresponding to
large $Q^2$ regime, they both show the peaks in the central $b$ and
a long tail in the region of large $b$. Conventionally, if we
consider the three quark core $| 3q\rangle$ in the proton or neutron,
the core is always located in the central $b$ and the size of the
core is expected to be smaller than $0.5~fm$. When the meson cloud,
pion meson cloud for example, is considered, the proton has the
components of $|(3q)^0\pi^+\rangle$ and $|(3q)^+\pi^0\rangle$ and the
neutron has the components of  $|(3q)^+\pi^-\rangle$ and $|
(3q)^0\pi^0\rangle$. Therefore, the long positive tail of the proton
transverse charge density comes from the charge pion cloud and on
the contrary, the long negative tail of the neutron transverse
density results from the negative pion cloud (see Fig. 8). Here for
the deuteron case (see Fig. 6), the long positive tail is expected
from the positive charged pion cloud \cite {Miller2}. The contribution from
the neutron negative pion cloud is less important and it is canceled by the
contribution of the proton. Moreover, the positive peak of the transverse
charge density at central $b$ is smaller than the proton, since the
contribution of the positive proton peak is partly canceled by the negative
peak of the neutron and moreover the loop integral also suppresses the peak.
So far, the origin of the negative peak of the
neutron transverse charge density is still an open question \cite {Miller2}. \\

\section{Summary}

To summarize this work, we use a phenomenological effective
Lagrangian approach to study the EM form factors of the deuteron,
particularly, the transverse charge and magnetic densities of the
deuteron. We show the EM form factors of the deuteron and their
transverse densities in the light-front representation. We find that
the present approach could reproduce the EM form factors, at least
qualitatively, although it is simple with only one parameter. The
important issue, in this work, is the study of the transverse
densities, particularly of  the transverse charge density of the
deuteron.  We find the transverse charge density reaches its maximum
at the central $b$ and it has a long positive tail. It means that the
charge quark core is located at the central $b$ and in the large $b$
region, the positive charge pion cloud dominates. This phenomenon is
similar to the proton case. Moreover our analysis shows the
remarkable differences between the two different parametrizations in
the central $b$ region. This is due to the fact that we know much
about the charge form factor of the deuteron in the low $Q^2$
region, but less in the large $Q^2$ region and the latter
one corresponds to the small $b$ region.\\

It should be reiterated that our present approach is simple and it
can be further improved. Here our estimated charge and magnetic form factors
fit the data qualitatively. We did not show the estimated quadrupole
form factor of the deuteron. This is  because the quadrupole form factor
is sensitive to the $D$-wave component of the deuteron, which we did not
include explicitly. A more sophisticated calculation with a more realistic
description of the deuteron wave function including $D$-wave component
is in progress. \\

\begin{acknowledgments}

This work is supported by National Sciences Foundations of China
Nos. 10975146, 11035006, 11261130 and 11165005, as well as supported, in part,
by the DFG and the NSFC through funds provided to the Sino-Germen CRC
110 ``Symmetries and the Emergence of Structure in QCD''. YBD also
thanks the Institute of Theoretical Physics, University of T\"ubingen
for the warm hospitality and thank the support from the Alexander von
Humboldt Foundation.

\end{acknowledgments}

\appendix

\section{The scalar loop integral}\label{LoopApp}
There are some discussions about the scalar loop integral as \cite {Melo,
Miller1}
\eq
I(P^2,q^2,P\cdot q)=\int \frac{d^4k}{(2\pi)^4}
\frac{1}{\big (k^2-M_N^2+i\epsilon\big )
\big [(k+q)^2-M_N^2+i\epsilon\big ]\big [(P-k)^2-M_N^2+i\epsilon\big ]}.
\en
In the light-front representation , $a^{\pm}=a^0\pm a^3$, and
$\int d^4k=\int \frac{1}{2}dk^+dk^-d^2\vec{k}$, then the
integral of $I(P^2,q^2,P\cdot q)$ is
\eq
I(P^2,q^2,P\cdot q)&=&\int \frac{dk^+dk^-d^2\vec{k}}{2(2\pi)^4}
\frac{1}{k^{+2}(P^+-k^+)}\\ \nonumber
&&\times\frac{1}{[k^--\frac{\vec{k}^2+M_N^2}{k^+}+\frac{i\epsilon}{k^+}]
\Big [k^--\frac{(\vec{k}+\vec{q})^2+M_N^2}{k^+}+\frac{i\epsilon}{k^+}\Big ]
\Big [P^--k^--\frac{(\vec{P}-\vec{k})^2+M_N^2}{P^+-k^+}+\frac{i\epsilon}
{P^+-k^+}\Big ]}.
\en
One may integrate over the upper half plane of the complex $k^-$, as show
in Fig. 9, and one can find a non-vanishing contribution only for the case
\begin{figure}
\centering
\includegraphics[width=10cm]{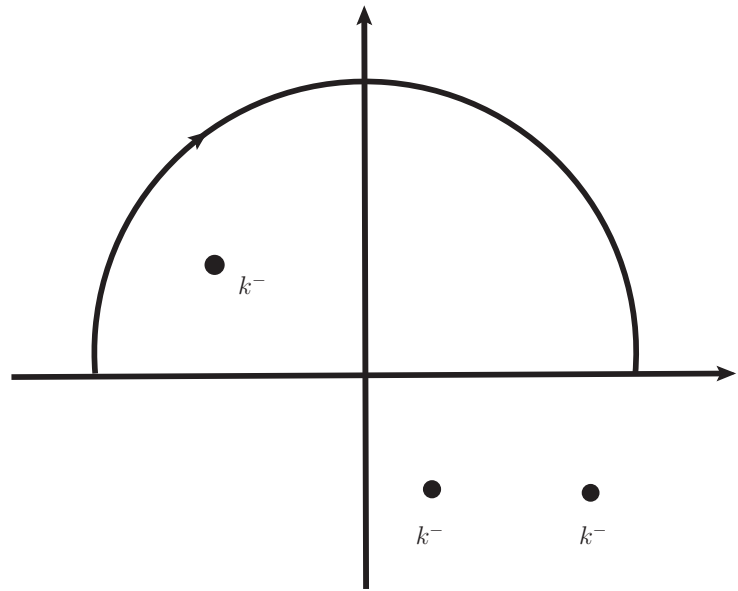}\\
\caption{The $k^-$ complex plane.}
\end{figure}
of $0< k^+< P^+$. Then
\eq
I(P^2,q^2,P\cdot q)=-i\int \frac{d^2\vec{k}}{2(2\pi)^3}\int
\frac{dk^+}{k^{+2}(P^+-k^+)}
\frac{1}{P^--\frac{\vec{k}^2+M_N^2}{k^+}
-\frac{(\vec{P}-\vec{k})^2+M_N^2}{P^+-k^+}}
\frac{1}{P^--\frac{(\vec{k}+\vec{q})^2+M_N^2}{k^+}
-\frac{(\vec{P}-\vec{k})^2+M_N^2}{P^+-k^+}},
\en
whereas, $k^+<0$ and $k^+>P^+$ doesn't contribute to the integral.\\

Define $x=\frac{k^+}{P^+}$, and choosing the reference frame of
$q^+=q^-=0$ \cite {Melo, Miller1},
then the above equation can be expressed as
\eq
I(P^2,q^2,P\cdot q)=-i\int \frac{d^2\vec{k}}{2(2\pi)^3}\int_0^1
\frac{dx}{x(1-x)}
\frac{1}{P^+P^--\frac{\vec{k}^2+M_N^2}{x}
-\frac{(\vec{P}-\vec{k})^2+M_N^2}{1-x}}
\frac{1}{P^+P^--\frac{(\vec{k}+\vec{q})^2+M_N^2}{x}
-\frac{(\vec{P}-\vec{k})^2+M_N^2}{1-x}}
\en

\end{document}